\documentclass[modern]{aastex62}
\usepackage{natbib}
\usepackage{placeins}
\usepackage{rotfloat}
\usepackage{gensymb}

\makeatletter
\def \blfootnote{\xdef\@thefnmark{}\@footnotetext}
\makeatother

\begin{document}
\title{Thermal Emission in the Southwest Clump of VY~CMa\footnote{The LBT is an international collaboration among institutions in the United States, Italy and Germany. LBT Corporation partners are: The University of Arizona on behalf of the Arizona Board of Regents; Istituto Nazionale di Astrofisica, Italy; LBT Beteiligungsgesellschaft, Germany, representing the Max-Planck Society, The Leibniz Institute for Astrophysics Potsdam, and Heidelberg University; The Ohio State University, and The Research Corporation, on behalf of The University of Notre Dame, University of Minnesota and University of Virginia.}}

\author[0000-0002-1913-2682]{Michael S.\ Gordon}
\affiliation{Minnesota Institute for Astrophysics, School of Physics and Astronomy\\116 Church St SE, University of Minnesota, Minneapolis, MN 55455, USA}
\affiliation{SOFIA Science Center, NASA Ames Research Center, Moffett Field, CA 94035, USA}

\author[0000-0002-8716-6980]{Terry J.\ Jones}
\affiliation{Minnesota Institute for Astrophysics, School of Physics and Astronomy\\116 Church St SE, University of Minnesota, Minneapolis, MN 55455, USA}

\author[0000-0003-1720-9807]{Roberta M.\ Humphreys}
\affiliation{Minnesota Institute for Astrophysics, School of Physics and Astronomy\\116 Church St SE, University of Minnesota, Minneapolis, MN 55455, USA}

\author[0000-0002-2314-7289]{Steve Ertel}
\affiliation{Steward Observatory, Department of Astronomy, University of Arizona, 993 N. Cherry Ave, Tucson, AZ 85721, USA}

\author{Philip M.\ Hinz}
\affiliation{Steward Observatory, Department of Astronomy, University of Arizona, 993 N. Cherry Ave, Tucson, AZ 85721, USA}

\author{William F.\ Hoffmann}
\affiliation{Steward Observatory, Department of Astronomy, University of Arizona, 993 N. Cherry Ave, Tucson, AZ 85721, USA}

\author[0000-0003-0454-3718]{Jordan Stone}
\affiliation{Steward Observatory, Department of Astronomy, University of Arizona, 993 N. Cherry Ave, Tucson, AZ 85721, USA}

\author{Eckhart Spalding}
\affiliation{Steward Observatory, Department of Astronomy, University of Arizona, 993 N. Cherry Ave, Tucson, AZ 85721, USA}

\author{Amali Vaz}
\affiliation{Steward Observatory, Department of Astronomy, University of Arizona, 993 N. Cherry Ave, Tucson, AZ 85721, USA}

\correspondingauthor{Michael S.\ Gordon}
\email{gordon@astro.umn.edu}

\begin{abstract}
  We present high spatial resolution LBTI/NOMIC 9--12~\micron\ images of VY~CMa and its massive outflow feature, the Southwest (SW) Clump. Combined with high-resolution imaging from HST (0.4--1~\micron) and LBT/LMIRCam (1--5\micron), we isolate the spectral energy distribution (SED) of the clump from the star itself. Using radiative-transfer code \texttt{DUSTY}, we model both the scattered light from VY~CMa and the thermal emission from the dust in the clump to estimate the optical depth, mass, and temperature of the SW~Clump. The SW~Clump is optically thick at 8.9~\micron\ with a brightness temperature of $\sim$200~K. With a dust chemistry of equal parts silicates and metallic iron, as well as assumptions on grain size distribution, we estimate a dust mass of $5.4\times10^{-5}\,M_\odot$.  For a gas--to--dust ratio of 100, this implies a total mass of $5.4\times10^{-3}\,M_\odot$. Compared to the typical mass-loss rate of VY~CMa, the SW~Clump represents an extreme, localized mass-loss event from $\lesssim300$ years ago.
\end{abstract}

\keywords{stars: individual (\object[V* VY CMa]{VY~CMa}) --- stars: mass-loss --- stars: winds, outflows --- supergiants}

\section{Introduction}
The extreme red supergiant VY~Canis~Majoris is one of the brightest infrared sources in the sky. HST imaging and long-slit spectroscopy from 0.4 to 1~\micron\ reveal a complex circumstellar nebula environment with multiple arcs and knots \citep{smith2001,humphreys2005,humphreys2007}. Ejected in separate mass-loss events over the past $\sim$1000 years, these features are structurally and kinematically distinct from the surrounding nebulosity.

\cite{shenoy2013} extended the exploration of VY~CMa's ejecta into the near- to mid-infrared with higher spatial resolution than previous studies with ground-based 1--5~\micron, adaptive optics imaging using LMIRCam \citep{skrutskie2010} on the Large Binocular Telescope (LBT). The dominant IR source in the 2.2, 3.8, and 4.8~\micron\ ($\mathrm{K_s}$, L$^\prime$, and M band) images is the peculiar ``Southwest Clump'' (hereafter, SW~Clump), which is optically thick in the HST/WFPC2 images at 1~\micron\ \citep{smith2001}. \cite{shenoy2013} determined that the high surface brightness of the SW~Clump requires optically-thick scattering at wavelengths shorter than 5~\micron, rather than thermal emission from dust grains since the expected blackbody equilibrium temperature for material $\sim$1500~AU from the central star is quite low ($\lesssim170$~K).

Scattering as the dominant component of the SW~Clump has been confirmed using high-resolution imaging polarimetry in the near-IR. Using MMT-Pol \citep{packham2012} on the 6.5m MMT Observatory at Mt Hopkins, \cite{shenoy2015} observed $\sim$30\% fractional polarization in the clump at 3.1~\micron, which requires optically-thick scattering from low albedo dust grains. In earlier work, \cite{shenoy2013} estimate a lower-limit on the total mass within the clump of $0.5-2.5\times10^{-2}\,M_\odot$ depending on the assumed gas--to--dust ratio (see discussion in \S\ref{sec:mass}). In any case, this ejecta event can be contrasted with VY~CMa's ``normal'' mass-loss rate of $\sim$10$^{-4}\,M_\odot$~yr$^{-1}$ \citep{danchi1994,humphreys2005,decin2006}, suggesting that the SW~Clump represents a single mass-loss episode from a localized region of VY~CMa's stellar atmosphere.

Recent sub-millimeter observations with ALMA reveal dusty concentrations within $\sim$10 $R_\star$ of VY~CMa \citep{richards2014,ogorman2015,vlemmings2017}, adopting the \cite{wittkowski2012} measurement $R_\star = 1420\,R_\odot$. \cite{ogorman2015} found a cold clump to the southeast, ``Clump~C,'' located closer to VY~CMa than the SW~Clump---400~AU (61~$R_\star$) vs.\ 1500~AU (230~$R_\star$).  While \cite{ogorman2015} estimate a dust mass lower limit of $2.5\times10^{-4}\,M_\odot$ for Clump~C, similar to the SW~Clump, there is no evidence for the SW~Clump in the ALMA images at 321 and 658~GHz \citep[Bands 7 and 9;][]{ogorman2015} or at 178~GHz \citep[Band 5;][]{vlemmings2017} in thermal emission. \cite{kaminski2013} did not observe the SW~Clump in thermal emission with the Submillimeter Array (SMA), though it was observed in line maps of H$_2$S (300.5~GHz), CS (293.9~GHz), and in several other molecular transitions. Given the mass estimates of the SW~Clump from the LMIRCam and MMT-Pol observations in \cite{shenoy2013,shenoy2015}, the non-detection in thermal emission in the ALMA bands may have implications for the dust grain properties in the far-IR.

Even without detection of continuum emission of the SW~Clump in the radio, molecular transition studies from ALMA and the SMA are useful in tracing the geometry of the clump, particularly the clump's orientation relative to the plane of the sky. Slightly blue-shifted TiO$_2$ emission observed in ALMA observations \citep{debeck2015} appears co-incident with scattered light in the 1~\micron\ HST images \citep{smith2001} and suggests that the SW~Clump---or at least material between the clump and the star---is partially in front of the plane of sky \citep{debeck2015}. However, NaCl emission at the location of the SW~Clump appears redshifted at $\sim$3~km~s$^{-1}$ \citep{decin2016} with respect to the LSR velocity, consistent with the \cite{humphreys2007} kinematic study of HST images. For the analysis in this work, we assume, then, that the SW~Clump is at least close to the plane of the sky.

In this study, we present LBT/NOMIC \citep{hoffmann2014} 8.9, 10.3, and 11.9~\micron\ imaging and photometry of VY~CMa and its SW~Clump. While the earlier LMIRCam observations reveal the scattered light of the dusty clump, NOMIC imaging provides measurements of the thermal emission of the dusty grains. We model the spectral energy distributions of both VY~CMa and the SW~Clump separately using the radiative-transfer code \texttt{DUSTY} \citep{ivezic1997} to show that the thermal emission at 8--12~\micron\ is largely consistent with a non-detection by ALMA at 400--1000~\micron, but slightly above the ALMA detection limit at 1.7mm.

\section{Observations \& Data Reduction}\label{sec:obs}
\begin{figure}[h!t]
  %\epsscale{0.75}
  \plotone{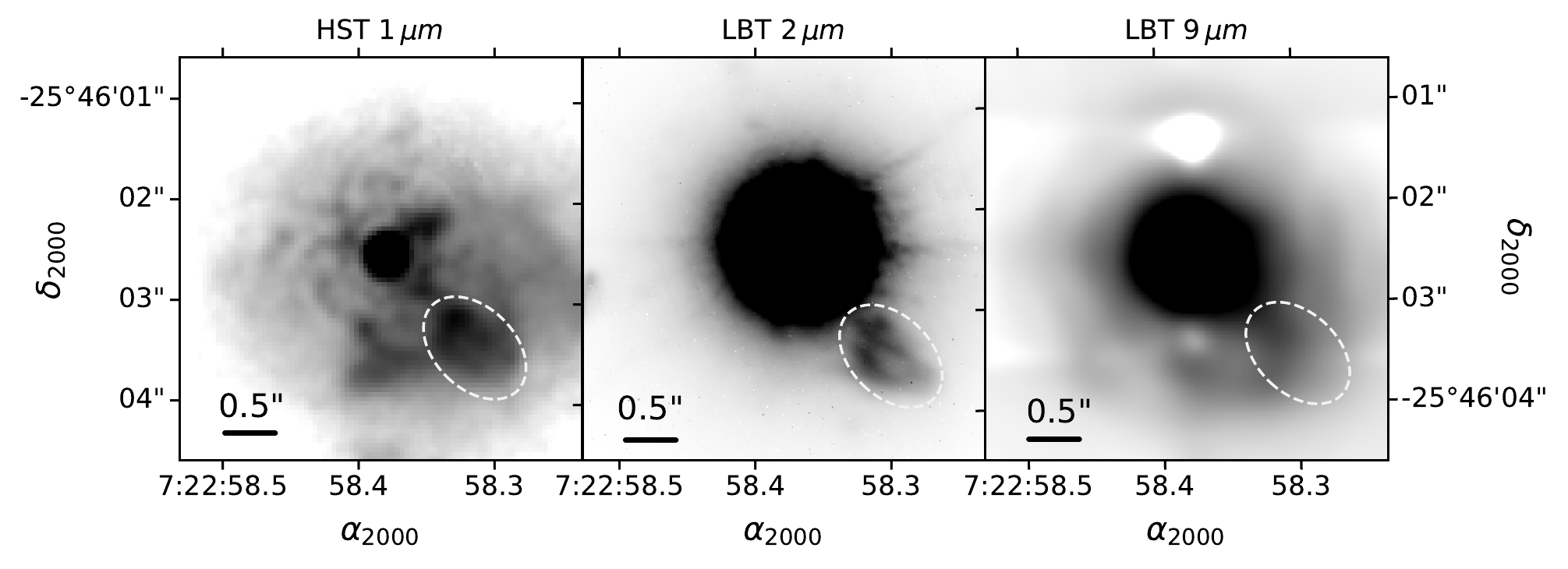}
  \caption{\mbox{\textit{Left:}} HST/WPFC2 1~\micron\ \citep[F1042M;][]{smith2001}. \mbox{\textit{Center:}} LBT/LMIRCam 2.2~\micron\ \citep[$\mathrm{K_s}$-band;][]{shenoy2015}. \mbox{\textit{Right:}} LBT/NOMIC 8.9~\micron\ (this work). The white bands in the NOMIC image are artifacts due to column saturation around the central star. The white dashed region represents the elliptical aperture from \cite{shenoy2013} defined as roughly 1~$\sigma$ above the background in the LMIRCam K$_s$ image. This region is $\sim0.6\times0.4\arcsec$, centered $\sim$1.5\arcsec\ from the star, and inclined 45\degree\ East from North.}
  \label{fig:montage}
\end{figure}

\begin{figure}[h!t]
  %\epsscale{0.75}
  \plotone{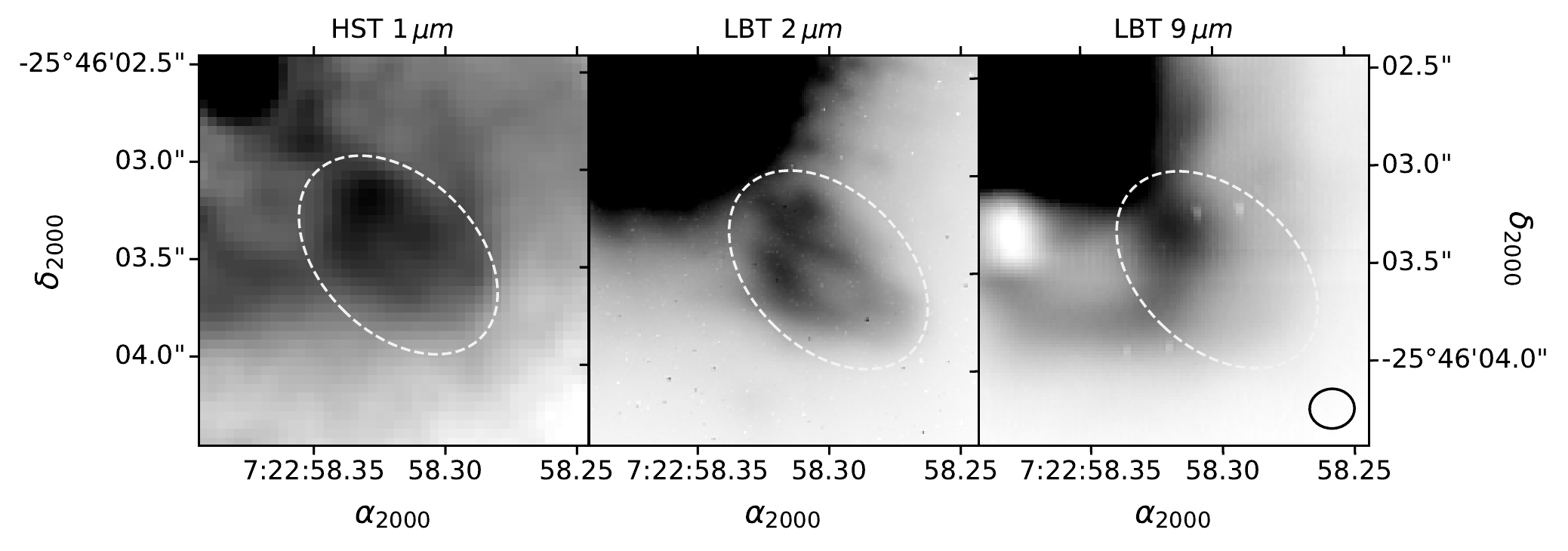}
  \caption{Same as in Figure~\ref{fig:montage} but zoomed in on the SW~Clump feature. The black ellipse in the bottom right corner is the FWHM of the PSF of the NOMIC 8.9~\micron\ image.}
  \label{fig:montagezoom}
\end{figure}

We observed VY~CMa with NOMIC on UT 2017 January 12 with a single 8.4~m primary mirror on the LBT. The Nulling Optimized Mid-Infrared Camera \citep[NOMIC;][]{hoffmann2014} is part of the Large Binocular Telescope Interferometer \citep[LBTI;][]{hinz2016} system. It uses a $1024\times1024$ Si:As array with a pixel scale of 0.018\arcsec\ pix$^{-1}$ and provides a field-of-view of $12\arcsec\times12\arcsec$. Images were made at 8.9~\micron\ ($\Delta\lambda=0.76\,\micron$), 10.3~\micron\ ($\Delta\lambda=6.0\,\micron$), and 11.9~\micron\ ($\Delta\lambda=1.13\,\micron$) with individual exposure times of 27.5~milliseconds for a total of $\sim$90~seconds in each filter ($\sim$3200 individual frames).  The exposure times were short to mitigate saturation from the central star, and the telescope was nodded between two positions on the NOMIC chip to ease background subtraction in data reduction. The reduced 8.9~\micron\ image is shown in Figure~\ref{fig:montage} on the right, aligned with the HST/WFPC2 1~\micron\ image from \cite{smith2001} and the LBT/LMIRCam $\mathrm{K_s}$-band image from \cite{shenoy2015}.  Figure~\ref{fig:montagezoom} shows the same three frames zoomed in on the SW~Clump.

\begin{figure}[h!t]
  %\epsscale{0.75}
  \plotone{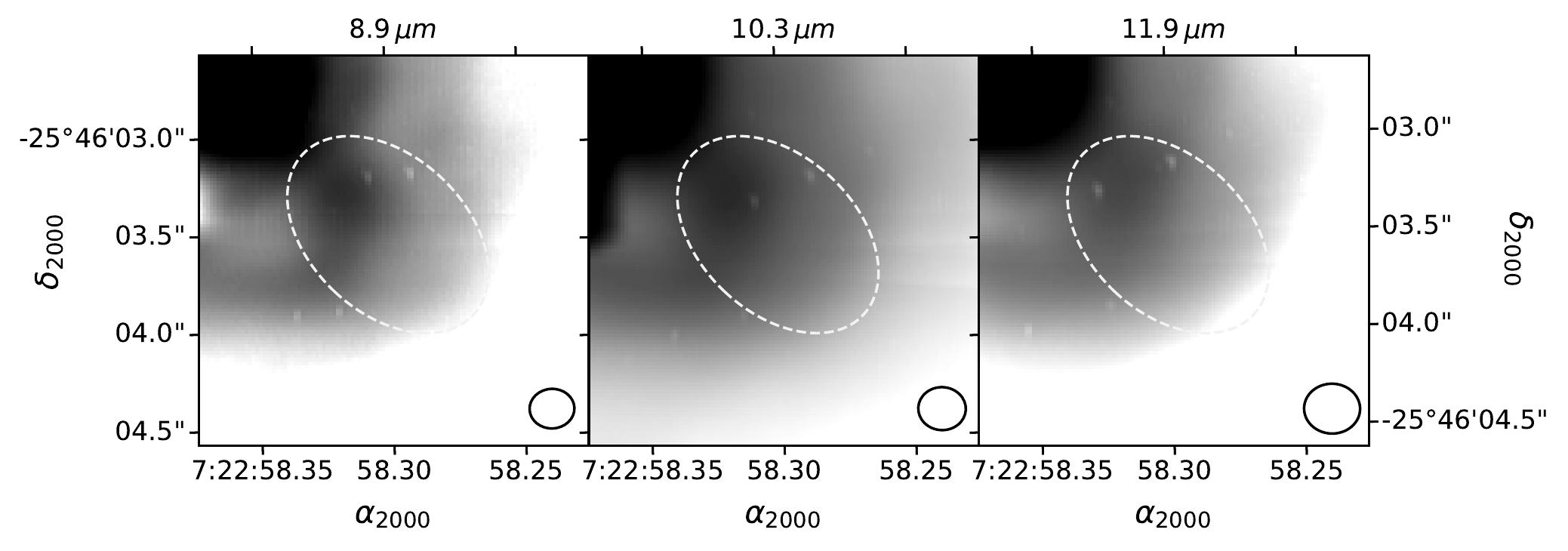}
  \caption{Observations in all three NOMIC filters zoomed of the SW~Clump feature. The observed PSF in each band is indicated in the bottom right of each panel. We have tried to scale these images to clearly display the clump, but the brightness of VY~CMa creates contrast issues.}
  \label{fig:montagezoomnomic}
\end{figure}

Sirius was observed at similar airmass and with the same nod locations on the NOMIC array for both flux and point-spread function (PSF) calibration. The PSFs were modeled in each wavelength at each nod position using the Astropy \citep{astropy2013} fitting functions for a two-dimensional Gaussian.  For flux calibration, we used photometry of Sirius with Gemini/T-ReCS from \cite{skemer2011}.  The T-ReCS and NOMIC filters have similar central wavelengths but different filter bandwidths, so we scale our measured counts into ``synthetic filters'' to effectively interpolate the Sirius photometry into the NOMIC filter sets.  This filter correction permits flux calibration of our VY~CMa images.

For each nod position, the $\sim$3200 frames in each filter are mean-combined with a sigma clipping threshold of three standard deviations from the average in each pixel.  The two nod position images are subtracted from each other, VY~CMa is masked out, and the background RMS in each NOMIC amplifier is modeled separately using the Astropy-affiliated \texttt{photutils}\footnote{\footnotesize{\texttt{photutils} provides tools for detecting and measuring the photometry of astronomical sources. The software is still in development, with documentation available at \url{https://photutils.readthedocs.io/}.}} package.

\section{Results \& Discussion}
\subsection{Photometry of the SW Clump}\label{sec:phot}

To quantify the flux in the SW~Clump relative to VY~CMa's SED, we need to subtract the contribution from the central star itself. In a manner similar to \cite{shenoy2013} we scale the amplitude of the PSF models from the Sirius images to match the profile of VY~CMa.  Since the central star is partly saturated, the ``wings'' of the PSF are used in the scaling to both locate the centroid and scale the amplitude.  While centroiding on a saturated source can be uncertain, at the distance of the SW~Clump from the star, the flux contribution from the PSF was minimal ($\lesssim10\%$ of the flux in the clump at 8.9~\micron). In all three bands, saturation from the central star causes column bleed artifiacts, but this saturation fortunately missed our aperture of interest.

We recalculate the SW~Clump photometry from \cite{shenoy2013} on the LMIRCam $\mathrm{K_s}$, L$^\prime$, and M band images for consistent treatment with the NOMIC images.  We generate an aperture around the SW~Clump using the $\mathrm{K_s}$ image to define a region which extends to $1\,\sigma$ above the background.  This aperture is roughly elliptical ($0.61\times0.44\arcsec$ beam) and centered $\sim$1.5\arcsec\ from the central star inclined at 45\degree\ East from North.  Photometry is performed with \texttt{photutils}, and the same aperture is used in all the LMIRCam and NOMIC images. Additionally, we apply this aperture to the HST/WFPC2, PSF-deconvolved images from \cite{smith2001} in the F410M, F547M, and F1042M medium-width continuum filters, and the narrow H$\alpha$ filter (F656N). In the HST optical images, several arcs, knots, and clumpy features are resolved within the large aperture, so the measured photometry is likely an overestimate.  Additionally, without radial velocity measurements of each of these resolved sub-clumps, we cannot determine which of these features are actually coincident with the SW~Clump mass--loss event.  However, the aperture photometry in the optical is performed in the same manner as for the IR images for consistency.  The SED models described in \S\ref{sec:dusty} do not weight the optical photometry to determine the best fit.

\begin{deluxetable}{llrrrrr}
  \tablecaption{Photometry of the SW~Clump\label{tab:phot}}
  \tablecolumns{7}
  \tablenum{1}
  \tablehead{\colhead{Telescope} & \colhead{Instrument} & \colhead{Date Obs} & \colhead{Filter} & \colhead{$\lambda_0$} & \colhead{Flux} & \colhead{Sys.\ Error\tablenotemark{$\dagger$}} \\
    \colhead{} & \colhead{} & \colhead{(UT)} & \colhead{} & \colhead{(\micron)} & \colhead{(Jy)} & \colhead{(Jy)}}
  \startdata
  HST & WFPC2 & 22 Mar 1999 & F410M & 0.4 & $6.6\times10^{-3}$ & $1.9\times10^{-3}$ \\
  HST & WFPC2 & 22 Mar 1999 & F547M & 0.5 & 0.1 & 0.04 \\
  HST & WFPC2 & 22 Mar 1999 & F656N & 0.7 & 0.4 & 0.15 \\
  HST & WFPC2 & 22 Mar 1999 & F1042M & 1.0 & 2.1 & 0.91 \\
  LBT & LMIRCam & 16 Nov 2011 & $\mathrm{K_s}$ & 2.2 & 6.8 & 1.2\phn \\
  LBT & LMIRCam & 16 Nov 2011 & L$^\prime$ & 3.8 & 15.0 & 4.1\phn \\
  LBT & LMIRCam & 16 Nov 2011 & M & 4.8 & 29.1 & 11.8\phn \\
  LBT & NOMIC & 12 Jan 2017 & 8.9 & 8.9 & 186.9 & 32.4\phn \\
  LBT & NOMIC & 12 Jan 2017 & 10.3 & 10.3 & 318.8 & 75.5\phn \\
  LBT & NOMIC & 12 Jan 2017 & 11.9 & 11.9 & 389.4 & 84.2\phn \\
  \hline
  ALMA & \nodata & 16 Aug 2013 & Band 9 & 456 & 0.75\tablenotemark{*} & \nodata \\
  ALMA & \nodata & 16 Aug 2013 & Band 7 & 934 & $1.6\times10^{-2}$\tablenotemark{*} & \nodata \\
  ALMA & \nodata & 16 Oct 2016 & Band 5 & 1680 & $8.1\times10^{-4}$\tablenotemark{*} & \nodata \\
  \enddata
  \tablenotetext{*}{Fluxes represent the 3-$\sigma$ upper limits estimated from the RMS noise in ALMA data scaled to the SW~Clump aperture size.}
  \tablenotetext{\dagger}{Systematic error reported as the standard deviation of the flux in a grid of apertures with different center positions (see discussion in text). Photometric error in the flux-calibrated NOMIC images is estimated at $<10\%$.}
\end{deluxetable}

The photometry is summarized in Table~\ref{tab:phot}. Since the SW~Clump is diffuse and we are uncertain of its total spatial extent, we generate a grid of apertures, all with the same total area, but allowing the center to move 0.1\arcsec\ in all directions.  The error value in Table~\ref{tab:phot} is the standard deviation of this aperture grid and represents here our measure of systematic uncertainty in the flux. Also included are flux limits for three ALMA bands. VY~CMa was observed as ALMA Science Verification data on UT 2013 August 16-19 \citep[321, 658~GHz;][]{richards2014,ogorman2015} and on UT 2016 October 16 \citep[178~GHz;][]{vlemmings2017}. As the continuum emission from the SW~Clump was undetected in these bands, we instead report a flux limit as $3\,\times$ the root-mean-square (RMS) noise in each image, where the measured RMS in the ALMA images is scaled to the beam-size of our photometric aperture. For example, with the synthesized ALMA beam at 178~GHz of $\sim0.5\times0.2\arcsec$ with an RMS noise of 0.1~mJy beam$^{-1}$ \citep{vlemmings2017} and our $0.61\times0.44\arcsec$ aperture beam ($2.7\,\times$ ALMA beam-size), then the detection limit assuming the total flux of the SW~Clump is distributed evenly over the beam would be 0.1~mJy beam$^{-1}\times2.7$ beams $\times\,3$ limit $\approx0.8$~mJy.  These limits are included in Table~\ref{tab:phot}.

\begin{figure}[h!t]
  %\epsscale{0.75}
  \plotone{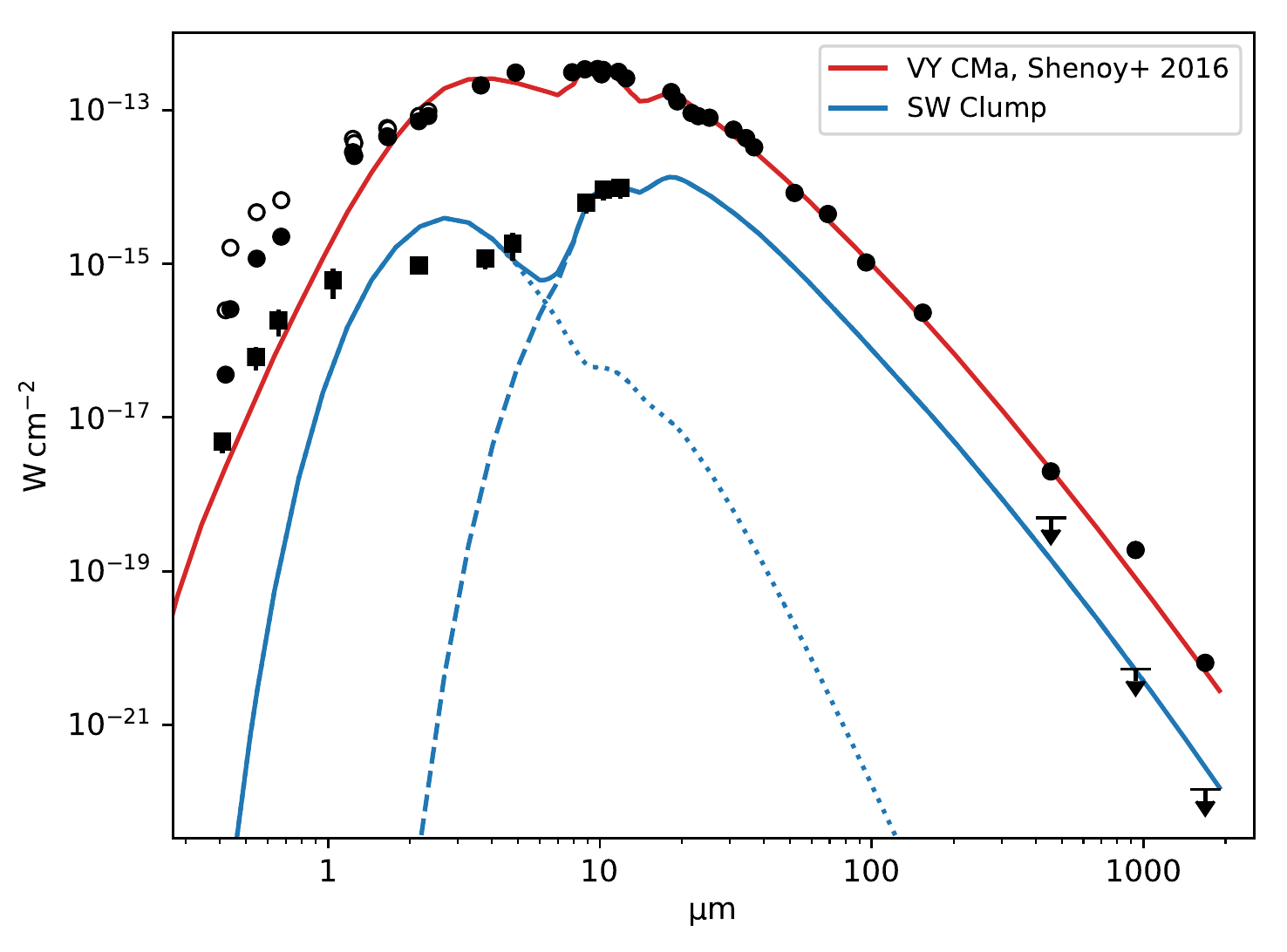}
  \caption{Photometry of VY~CMa and the SW~Clump.  Closed circles represent photometry of VY~CMa compiled from the literature. Open circles are the interstellar extinction-corrected optical and near-IR photometry.  Black squares are the photometry from this work on the SW~Clump from WFPC2 (0.4--1~\micron), LMIRCam (2--5~\micron), and NOMIC (9--12~\micron).  The $3\sigma$ ALMA detection limits are shown as downward arrows in the submillimeter to millimeter. Errorbars represent total uncertainty in the flux calibration, PSF-subtraction, and systematic error in the aperture. The red line represents the best-fitting \texttt{DUSTY} model from \cite{shenoy2016} using a spherical dust distribution with a density profile of $\rho\left(r\right)\propto r^{-1.5}$. The blue line is the best fit slab model (this work) to the SW~Clump. The dotted line is the scattered light component for the slab model, and the dashed line indicates thermal emission, which begins to dominate at $\sim$5~\micron\ to longer wavelengths.}
  \label{fig:SED}
\end{figure}

The observed SEDs of both VY~CMa and the SW~Clump are shown in Figure~\ref{fig:SED}.  The closed circles represent photometry of VY~CMa compiled from the literature, including the HST/WFPC2 observations at 0.4--1~\micron\ plus the ESO~3.6~m telescope 1--20~\micron\ IR photometry from \cite{smith2001}, and the 20--40~\micron\ SOFIA/FORCAST and 60--150~\micron\ Herschel/PACS photometry from \cite{shenoy2016}.\footnote{\footnotesize{PACS data obtained as part of the guaranteed time Mass-loss of Evolved StarS (MESS) key program \citep{groenewegen2011}.}} The open circles are the extinction-corrected optical and near-IR photometry for foreground (interstellar) A$_V = 1.5$ \citep{shenoy2015} and a traditional extinction curve \citep{cardelli1989}.  The black squares are the photometry from this work on the SW~Clump using the elliptical aperture region discussed above in the WFPC2, LMIRCam, and NOMIC images.  The $3\sigma$ ALMA detection limits are shown as downward arrows in the submillimeter to millimeter. The model SED for the clump is fainter than the ALMA limits in the submillimeter regime but slighly above longward of 1mm, not inconsistent with their non-detection. None the less, our model does put strong constraints on the ALMA results.  \cite{ogorman2015} suggest that dust properties are different at mm waves from those we used to model the mid-infrared wavelength regime.

\subsection{\texttt{DUSTY} modeling}\label{sec:dusty}
To study the thermal properties of the SW~Clump, we model the SEDs of both VY~CMa and the SW~Clump using the \texttt{DUSTY} radiative-transfer code \citep{ivezic1997}.  \texttt{DUSTY} solves the one-dimensional radiative-transfer equation for either a spherically-symmetric dust distribution around a central source or through a slab of dusty material. For modeling the SED of VY~CMa itself, we employ the spherical mode of \texttt{DUSTY} following previous work in \cite{shenoy2016}, which analyzed the mass-loss histories around hypergiant stars $\mu$~Cep, IRC~+10420, $\rho$~Cas, and VY~CMa.  \cite{shenoy2016} fit a variety of dust density distributions to each star, and they found that for VY~CMa, a density profile of $\rho\left(r\right)\propto r^{-1.5}$ best explained the mid-infrared emission in the star's SED.

For our spherical \texttt{DUSTY} model, we adopt this dust density distribution as well as the chemistry from \cite{shenoy2016}---a 50-50 mixture of astronomical silicates from \cite{draine1984} and metallic iron from \cite{harwit2001}.  We assume the grain radii follow an MRN size distribution $n\left(a\right)\propto a^{-3.5}$ \citep{mathis1977} with $a_{min} = 0.005\,\micron$ and $a_{max} = 0.5\,\micron$. With an effective temperature of 3490~K \citep{wittkowski2012} and an assumed dust condensation temperature of 1000~K, \texttt{DUSTY} generates the model SED shown at the top of Figure~\ref{fig:SED} in red. While the dust condensation temperature can be modeled as a free parameter in \texttt{DUSTY} \citep[][varied $T_{in}$ in their models in the range of 500--1200~K]{beasor2016}, we assume a constant temperature of 1000~K to both reduce the dimensionality of our model sets and for consistency with previous work on RSG modeling in \cite{gordon2018} and see,  for example, \cite{groenewegen2012}.

\begin{figure}[h!t]
  %\epsscale{0.75}
  \plotone{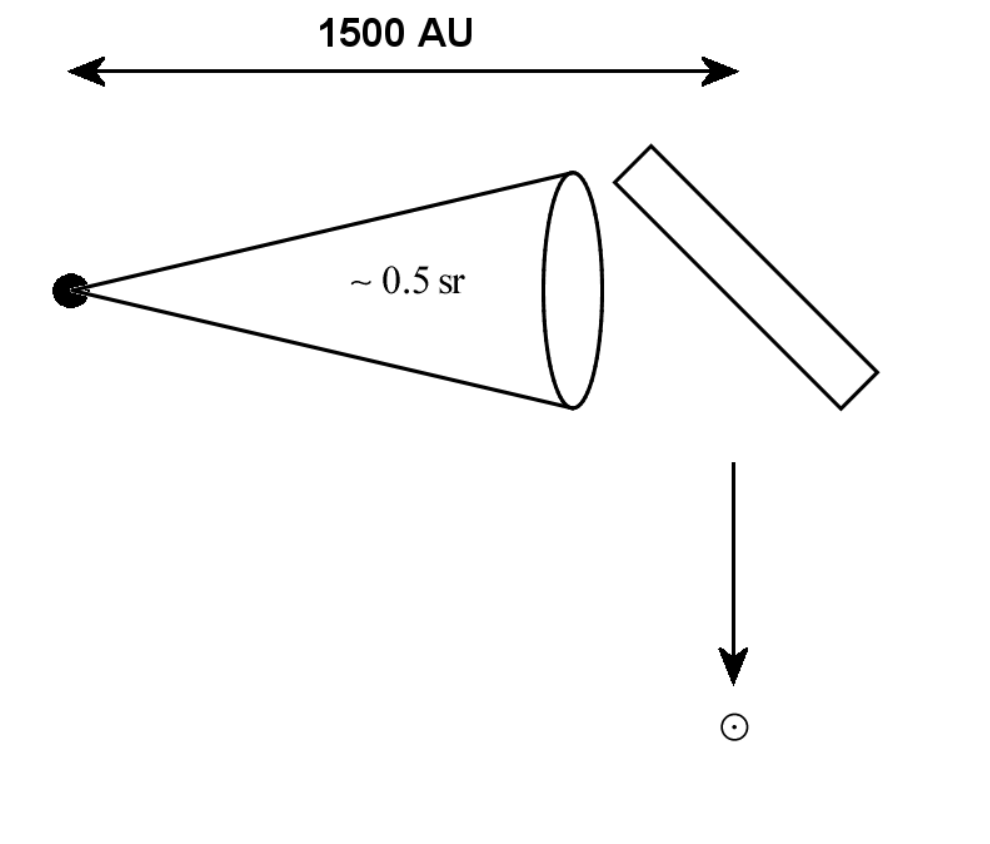}
  \caption{Geometry of the central star and the SW~Clump modeled as a slab in the \texttt{DUSTY} code.}
  \label{fig:geo}
\end{figure}

  \cite{shenoy2015} found that the SW~Clump was optically thick to scattering but also highly polarized with a fractional polarization of at least 30\%. Since optically-thick scattering tends to reduce the net polarization due to multiple scatters, the SW~Clump must be relatively close to the plane of the sky and the grains must have an albedo $\omega \la 0.4$ to achieve the measured level of polarization. Rather than model the clump separately in scattered light and emitted light, we use the ``slab'' mode in \texttt{DUSTY} for the SW~Clump, which reproduces the scattered (reflected) and thermal emission from a central source on some planar geometry. Figure~\ref{fig:geo} illustrates the geometry of our experimental setup. The actual 3D morphology of the SW~Clump is unknown, so the use of a simple slab is clearly an approximation. A spherical geometry for the clump is likely more consistent with the polarimetry; however, \texttt{DUSTY} can not model this case.

To reproduce the SW~Clump emission, we generate a grid of models varying the optical depths of the SW~Clump material at 8.9~\micron\ ($0.01 < \tau_{8.9} < 5$). Since the SW~Clump is located within a circumstellar nebula of dusty material, the radiation incident on the clump will include light partially extinguished from the star as well as radiation from hot dust between the star and the clump. Examination of the SED indicates that the bulk of the hot dust emission interior to the SW clump is emitted between 1 and 5~\micron.  Therefore, we approximate the central source seen from the clump as a blackbody with effective temperature between 1000 and 2000~K while maintaining the total bolometric flux.  A blackbody in this temperature range would roughly peak between 2 and 5~\micron. Since there is no spatial information on the dust emission close to the star producing the bulk of the $2-5~\micron$ emission, we did not use \texttt{DUSTY} to model a compact shell with an outer edge at the projected distance of the SW~Clump, which would have required an unrealistic density profile.

To select a best-fitting model, we evaluate a reduced $\chi^2$ measurement of the observed SW~Clump photometry and the \texttt{DUSTY} output spectrum. Unlike for the spherical case, we scale the slab \texttt{DUSTY} model SEDs by the solid angle subtended relative to the central star, which for our elliptical aperture is $\sim$0.5~sr. The best-fitting model is shown in Figure~\ref{fig:SED} with the \texttt{DUSTY} input/output parameters summarized in Table~\ref{tab:models}. Our model fitting demonstrates that an optical depth around unity at 8.9~\micron\ is required for the slab to emit the observed flux in the mid-IR. This shows that the SW~Clump is optically thick to both scattering and emission, which means we can only observe the ``surface'' of the clump along the light-of-sight.  Thus, only a lower limit to the mass of the clump can be derived. The goodness-of-fit does not weight the HST (0.4--1~\micron) photometry here as we instead focus on the scattered and thermal emission present in the LMIRCam and NOMIC images for this work.

\begin{deluxetable}{lcc|ccc|ccc}
  \tablecaption{DUSTY Model Parameters \& Observed Temperatures (K)\label{tab:models}}
  \tablecolumns{9}
  \tablenum{2}
  \tablehead{
    \multicolumn{3}{c|}{Inputs} & \multicolumn{3}{c|}{Outputs} & \multicolumn{3}{c}{Observed} \\
    \hline
    \colhead{Model} & \colhead{$\mathrm{T_{eff}}$\tablenotemark{a}} & \colhead{$\tau_{8.9}$}\vline & \colhead{$\mathrm{f_{sc,\,5\mu m}}$\tablenotemark{b}} & \colhead{$\mathrm{T_{d}}$\tablenotemark{c}} & \colhead{$\mathrm{T_{color}}$\tablenotemark{d}}\vline & \colhead{$\mathrm{T_{BB}}$\tablenotemark{e}} & \colhead{$\mathrm{T_{bright}}$\tablenotemark{f}} & \colhead{$\mathrm{T_{color}}$\tablenotemark{g}}}
  \startdata
  \hline
  slab & 1600 & 1.03 & 92\% & 207 & 277 & 165 & 205 & 275 \\
  \enddata
  \tablenotetext{\scriptsize \textrm{a}}{Effective temperature of the input blackbody to model a ``pseudo-photosphere'' interior to the SW~Clump.}
  \tablenotetext{\scriptsize \textrm{b}}{Fractional contribution of scattered light to total SED flux at 5~\micron\ (M-band).}
  \tablenotetext{\scriptsize \textrm{c}}{Dust temperature in the SW~Clump measured from the model at the slab boundary facing the star. See \S\ref{sec:thermal} for discussion of the various temperature quantities.}
    \tablenotetext{\scriptsize \textrm{d}}{Color temperature from the model SED calculated from a ratio of the 8.9 and 11.9~\micron\ flux.}
  \tablenotetext{\scriptsize \textrm{e}}{Blackbody equilibrium temperature for a 270,000 $L_\odot$ central source \citep{wittkowski2012} at the distance of the SW~Clump.}
  \tablenotetext{\scriptsize \textrm{f}}{Brightness temperature for the measured flux in the SW~Clump.}
  \tablenotetext{\scriptsize \textrm{g}}{Color temperature in the SW~Clump aperture calculated from the 8.9 and 11.9~\micron\ images.}
\end{deluxetable}

The luminosity of the SW~Clump relative to the SED of VY~CMa itself serves as an independent check on the aperture area we derived from the 2~\micron\ LMIRCam. The bolometric flux of the clump, estimated by integrating the model curve from 0.3~\micron\ through 1mm, is about 3\% of the total luminosity of VY~CMa. Our clump aperture subtends a solid angle of $\sim$0.5~sr relative to the star, which is $\sim$4\% of the full sphere. Thus, our aperture area is consistent with the observed photometry.

We note here two of our greatest uncertainties in constraining the \texttt{DUSTY} models: the SED of circumstellar material between the star and the SW~Clump and the geometry of the SW~Clump. As discussed above, the SW~Clump is not illuminated by the 3490~K photosphere from VY~CMa, but rather a combination of attenuated light from the central star and emission from hot dusty material between the star and the clump.  We have provided as input to the \texttt{DUSTY} slab a simple blackbody with T$=1600$~K to approximate this incident SED, but the actual SED incident on the slab will certainly be more complicated. We note that a few hundred degree variation in the input blackbody temperature does not significantly alter the shape of the model SEDs from 5~\micron\ out to longer wavelengths.

The actual extent of the SW~Clump is not fully resolved in the NOMIC images.  The photometric aperture was defined from the LMIRCam K$_s$ image for consistency with \cite{shenoy2013}, but as we see in Figures~\ref{fig:montage} and \ref{fig:montagezoom}, the shape of the SW~Clump in scattered emission at 2~\micron\ is not the same as in thermal emission at 8.9~\micron. Additionally, \texttt{DUSTY} assumes isotropic scattering from dust grains without consideration of a dependence of the scattering efficiency on scattering angle. This may in part explain the different spectral shape observed from 1--5~\micron\ in Figure~\ref{fig:SED} in comparison to the model. Finally, as discussed above, the 3D morphology of the clump is unknown. The assumptions made for the geometry lead to uncertainty in the solid angle subtended by the SW~Clump relative to the central star; though, as described above, the fraction of the total sphere subtended by our aperture is consistent with the fraction of flux in the SW~Clump relative to VY~CMa's SED.

The data points for the SW~Clump at wavelengths shorter than $2.2~\micron$ lie significantly above the model in Figure~\ref{fig:SED}. It is possible that light directly from the photosphere of VY CMa is irradiating some dust along the line of sight, increasing the scattered flux. This radiation source is not in our model, since we are only interested in the scattered light from $2-5~\micron$. Also, dust in front of the SW~Clump that is optically thin at wavelengths longer than $2.2~\micron$, but scatter extra flux into the beam at shorter wavelengths, could also contribute to the discrepancy.

\subsection{Scattered vs.\ Thermal Emission}\label{sec:thermal}
\cite{shenoy2013} used the \texttt{BHMIE} code \citep{bohren1983} to calculate the extinction and scattering efficiencies of dust grains in the SW~Clump using Mie theory to determine the fractional contribution of scattering and thermal emission in the SED.  At 5~\micron, \cite{shenoy2013} estimates that $\sim$75\% of the flux in the SW~Clump is due to scattered light from VY~CMa. We can make a similar calculation since \texttt{DUSTY} also separates the scattered and thermal components of the SEDs, shown in Figure~\ref{fig:SED} with dotted and dashed lines, respectively.  We derive a fractional contribution from scattering at 5~\micron\ of 92\%. At wavelengths longer than 10~\micron, the emission is purely thermal.

In Table~\ref{tab:models}, we present several distinct temperature measurements. \texttt{DUSTY} provides estimates on the dust temperature as a function of optical depth through the slab.  For the surface of the slab facing the star, the dust temperature $\mathrm{T_d}$ is 207~K. As an independent check on consistency, we can roughly measure the dust temperature directly from the flux-calibrated NOMIC images. The analagous quantity to the \texttt{DUSTY} temperature at the slab's surface would be an observed brightness temperature in the IR, calculated for a blackbody as:
\begin{equation}
  \mathrm{T_{bright}} = \frac{hc}{k\lambda}\ln^{-1}\left(1+\frac{2hc^2}{I_\lambda \lambda^5}\right)
\end{equation}
For the total flux in our SW~Clump aperture at 8.9~\micron, this yields $\mathrm{T_{bright}} = 205$~K, similar to the ``physical'' dust temperature from \texttt{DUSTY} at the clump's surface.

We can also measure a color temperature from both the observed photometry and the model SED.  The NOMIC 8.9 and 11.9~\micron\ filters bracket the 10~\micron\ silicate emission feature, and therefore sample the continuum emission from dust in the SW~Clump. The ratio of the observed photometry yields a color temperature of 275~K. We recover a similar measurement from the model SED of 277~K, which is not surprising since the $\chi^2$-fitting performed on our grid of \texttt{DUSTY} models guarantees a recovered SED with a similar spectral shape to the observed photometry in the IR.  Also included for comparison in Table~\ref{tab:models} is the blackbody-equilibrium temperature evaluated at a distance of 1500~AU from a source with the bolometric luminosity of VY~CMa \citep[270,000 $L_\odot$;][]{wittkowski2012}.

\subsection{Mass Estimates}\label{sec:mass}
Since the optically-thick $\tau_{8.9}=1.03$ \texttt{DUSTY} model recovers the observed total flux in the SED, we can estimate the total mass in the SW~Clump from its optical depth, the grain size distribution, and the extinction efficiency.  Optical depth is defined as:
\begin{equation}\label{eq:tau}
  \tau_\lambda\equiv\int_{a_{min}}^{a_{max}}Q_\lambda\,n(a)\,\pi a^2\,da
\end{equation}
where $Q_\lambda$ is the extinction efficiency factor, and $n(a)$ is the column deinsity of grains with an MRN grain size distribution discussed in \S\ref{sec:dusty}. Since our input grains to the \texttt{DUSTY} models are 50\% silicates and 50\% iron, our efficiency is simply the average of the efficiency functions from \cite{draine1984} and \cite{harwit2001} (here, $Q\approx0.05$ at 8.9~\micron).  We assume both the extinction efficiency and the internal mass density of the grains ($\rho$) are constant with grain size which is reasonable since we are in the Rayleigh regime at $8-9~\micron$, and we define the column mass density as:
\begin{equation}\label{eq:density}
  m\left( {\rm{g}\,{\rm{c}}{{\rm{m}}^{{\rm{ - 2}}}}} \right)=\rho\int_{a_{min}}^{a_{max}}n(a)\,\frac{4}{3}\pi a^3\,da
\end{equation}
where $\rho$ is a typical grain mass density of 3~g~cm$^{-3}$. We find $m=4.2 \times 10^{-4}\left( {\rm{g}\,{\rm{c}}{{\rm{m}}^{{\rm{ - 2}}}}} \right)$. Multiplying by the total area in the clump---$2.6\times10^{32}$~cm$^2$ for our aperture at a distance of 1.2~kpc \citep[from VLBA parallax;][]{zhang2012}---we derive a mass of $5.4\times10^{-5}\,M_\odot$ in dust.  Adopting a gas:dust ratio of 100:1 \citep[for consistency with][]{shenoy2013}, yields a total mass (gas+dust) in the SW~Clump of $5.4\times10^{-3}\,M_\odot$. Given that the SW clump emits an amount of flux close to the fraction of flux from the star it intercepts, this mass must be considered a well constrained lower limit, given our assumptions regarding the grain population. 

This result is consistent with the lower limit estimate of $M\gtrsim5\times10^{-3}\,M_\odot$ from imaging polarimetry at 3.1~\micron\ by \cite{shenoy2013}.  If, however, we adopt the higher gas:dust ratio of 500:1 from \cite{decin2006} for VY~CMa, our mass estimate for the SW~Clump becomes $2.7\times10^{-2}\,M_\odot$.  Compared to the typical mass-loss rate of VY~CMa of $\sim$10$^{-4}\,M_\odot$~yr$^{-1}$ \citep{danchi1994}, such a large mass in a discrete feature likely represents an extreme, localized mass-loss event.

For comparison, \cite{ogorman2015} estimates a dust mass for the Clump~C feature of $\sim2\times10^{-4}\,M_\odot$, which they cite as a lower limit since their calculation is in the optically-thin regime. With additional Band~5 (178~GHz) ALMA data, \cite{vlemmings2017} updates this dust mass to $>1.2\times10^{-3}\,M_\odot$.  Clump~C is then almost two orders of magnitude more massive than the SW~Clump.  \cite{richards2014} and \cite{ogorman2015} also identified a second radio-bright continuum source at or near the center of the star that they call the VY component. This source, which is too close to the star for us to image, has a dust mass estimate of $\sim3\times10^{-5}\,M_\odot$, which is about half of our dust mass estimate for the SW~Clump. Relative to typical RSG mass-loss rates, these localized episodes of dusty ejecta are all extraordinary examples of the extreme outflow activity from massive evolved stars.

\section{Conclusions}
High-resolution, sub-arcsecond imaging from $9-12~\micron$ with NOMIC has allowed us to isolate the peculiar SW~Clump feature from the overall IR emission of VY~CMa. The resulting SED of the clump alone is a powerful tool in characterizing the thermal properties of the clump relative to the central star.  Through \texttt{DUSTY} modeling, we confirm that the clump is optically-thick from 9--12~\micron\ and has a brightness temperature of $\sim$200K.  With a firm lower limit to the dust mass of $5.4\times10^{-5}\,M_\odot$, the SW~Clump is comparable in mass to the radio-bright Clump~C and ``VY'' component identified in \cite{richards2014} and \cite{ogorman2015}.

At a distance of $\sim$1500~AU, the SW~Clump represents a recent mass-loss event from VY~CMa.  If we assume a  value for the velocity of 25~km~s$^{-1}$, typical for red supergiants, then the clump would have been ejected $\lesssim300$ years ago. 

Finally, we note that our models and estimates on thermal emission from the dust in the SW~Clump are not inconsistent with the non-detection at ALMA, but they do put strong constraints on the ALMA results. The SED models predict sub-mm fluxes at or below the 3$\sigma$ ALMA detection limits, but slightly above at 1.7mm using our SW~Clump aperture.  Our models are not at all constrained beyond the 11.9~\micron\ NOMIC photometry.  Therefore, high-resolution imaging of the SW~Clump in the $\sim$20--100~\micron\ regime is required to characterize fully the thermal emission from this fascinating mass-loss event.
\\
\\
M Gordon acknowledges support from the University of Minnesota Graduate School's Doctoral Dissertation Fellowship.

\facilities{LBT, LBTI (LMIRCam, NOMIC), HST (WFPC2)}

\software{Astropy \citep{astropy2013}, DUSTY \citep{ivezic1997}}

\end{document}